\documentclass[twocolumn]{aastex631}

\newcommand{\nustar}{\textit{NuSTAR}}
\newcommand{\ms}{\ensuremath{M_{\odot}}}

\newcommand{\lumcgs}{\ensuremath{\mathrm{erg}\,\mathrm{s}^{-1}}}

\usepackage{amsmath}

\shorttitle{Synchrotron cutoff in Ultraluminous X-ray sources}
\shortauthors{Ghosh et al.}
\graphicspath{{./}{figures/}}

\begin{document}

\title{Synchrotron cutoff in Ultraluminous X-ray sources}

\correspondingauthor{Tanuman Ghosh}
\email{tanuman@rri.res.in}

\author{Tanuman Ghosh}
\affiliation{Astronomy and Astrophysics, Raman Research Institute, C. V. Raman Avenue, Sadashivanagar, Bangalore 560080, India}

\author{Shiv Sethi}
\affiliation{Astronomy and Astrophysics, Raman Research Institute, C. V. Raman Avenue, Sadashivanagar, Bangalore 560080, India}

\author{Vikram Rana}
\affiliation{Astronomy and Astrophysics, Raman Research Institute, C. V. Raman Avenue, Sadashivanagar, Bangalore 560080, India}



\begin{abstract}
The origin of spectral curvature at energies $E\simeq 10$~keV in ultraluminous X-ray sources is not well understood. 
In this paper, we propose a novel mechanism based on synchrotron radiation to explain this cutoff. We show that relativistic plasma can give rise to observed spectral curvature for neutron star magnetic fields due to the variation in the latitude of synchrotron radiation. We analyze the \nustar\ data of two bright pulsar ULXs, NGC 5907 ULX1 and NGC 7793 P13, and provide estimates of the physical parameters of these sources. We fit the data for synchrotron emission at various latitudes and show that the spectral cutoff in these cases can be explained for a large range of acceptable physical parameters, e.g., a semi-relativistic plasma with $\gamma \simeq 20$ for high latitudes or a highly relativistic plasma ($\gamma \simeq 10^5$) for emission close to the electron's orbital plane in a typical magnetic field of $B\simeq 10^{12} \, \rm G$. We also discuss how such an emission mechanism can be distinguished from other proposed models. A corollary to our study is that most ULXs might be neutron stars as they display such a spectral cutoff.

\end{abstract}

\keywords{Radiative processes(2055) --- Neutron stars(1108) --- Ultraluminous x-ray sources(2164) --- High energy astrophysics(739)}


\section{Introduction} \label{sec:intro}

Ultraluminous X-ray sources (ULXs) are some of the brightest known  X-ray sources ($\rm L_x > 10^{39} \lumcgs$). Their luminosities exceed the classical Eddington limit of a $10$ \ms black hole (see \citealt{Kaaret2017} for a recent review). In addition, many  ULXs display a unique spectral curvature at energies $\simeq 10$~keV as shown  by broadband X-ray data (e.g., \citealt{Bachetti2013, Walton2013}). This spectral feature is one of the distinctive characteristics of ULXs compared to the hard state of Galactic X-ray binaries (XRBs) and active galactic nuclei (AGNs). The discovery of a neutron star ULX \citep{Bachetti2014} changed the perception of these sources, and a foremost conjecture is that a large fraction of the ULX population is neutron stars (e.g., \citealt{King2016, King2017, King2020}). Many theoretical models have been studied to explain emission mechanisms that generate such high luminosity from neutron stars (e.g., \citealt{Mushtukov2015, Mushtukov2017, Mushtukov2018, Mushtukov2019}). The origin of spectral cutoff, however, has no compelling theoretical model. Recent observational studies provide phenomenological models which invoke physical scenarios like compton scattering in the coronal region in low magnetic sources like black holes or comptonization in the accretion column in highly magnetized neutron stars (see e.g., \citealt{West2018, Walton2018apr, Walton2020}). In this paper, we propose an alternative model based on synchrotron radiation from different latitudes to explain the observed spectral cutoff. We explore the possible physical scenarios of this phenomenon in the context of ULXs and estimate physical parameters related to both the luminosity and the spectral cutoff in ULXs.

Synchrotron radiation is one of the most prevalent radiative processes in astrophysics \citep{Ribicky}. While the non-relativistic synchrotron radiation, the cyclotron radiation, provides a discrete spectrum, the emission by relativistic particles yields near-continuum spectrum owing to the higher harmonics contributing more predominantly to the observed spectrum (see \citealt{Landau,Ribicky} for a review). The astrophysical implications of synchrotron radiation are well studied in multiple wavelengths, including soft to hard X-rays (e.g., \citealt{Longair, Heinz2004, Maccarone2005, Markoff2005, Kisaka, Kisaka2017, Reigler}). In this paper, we explore the impact of high-latitude, optically-thin, classical  synchrotron emission on the  spectrum of the radiation for a range of speeds  encompassing a broad range from semi-relativistic  to ultra-relativistic electrons.

In the next section, we briefly review the physics of synchrotron radiation relevant to our work. We also provide  
approximate analytical expressions that allow one to study the emission from semi-relativistic to highly relativistic electrons for a range of latitudes. In \S~3, we provide details of the data we use and its pre-processing. The main results are presented in \S~4. In \S~5, we summarize our findings and discuss how our proposed method can be distinguished from other models.

\section{Synchrotron radiation: semi-relativistic to ultra relativistic transition}\label{sec:Theory}
We assume a geometric construct in which the incoherent synchrotron radiation originates close to the surface of a neutron star. As the length scale of magnetic fields is much larger than the curvature of the gravitating body, we can assume that the magnetic field lines are straight on scales from which the observed synchrotron emission occurs. Without loss of generality, we assume that the magnetic field is in the $z$-direction of cartesian geometry and the charged particles move in a circular motion around the uniform magnetic field lines in the xy plane.

The angular distribution of the radiated power in $n$th harmonic (or an angular frequency of observation, $\omega$) for a single electron   ($\rm erg \, sec^{-1}$)  per unit solid angle ($d\Omega$) can be expressed as \citep{Landau}: 
\begin{equation}
  dI_n = \frac{e^2\omega^2}{2\pi c} \left[\tan^2\theta J^2_n(n\beta\cos\theta)  +\beta^2 J'^2_n(n\beta\cos\theta) \right] {d\Omega}
  \label{eq:synrad_ang}
\end{equation}
Here $\beta = v/c$,   $B$ is the magnetic field strength, and $\theta$ is the angle between radiated emission and the particle's orbital plane. $J_n(x)$ is the Bessel function  and $J_n'(x)$ is  its derivative. The integer  $n$ denotes the discrete energy levels of electron's energy with  $\omega = n \omega_B$. $\omega_B = eB/\gamma m_e c$,    $\gamma = 1/\sqrt{1-\beta^2}$ is the relativistic boost.

The aim of our study is to analyze   the  emission from both semi-relativistic and  ultra-relativistic plasmas. Eq.~(\ref{eq:synrad_ang}) allows for the transition from cyclotron to synchrotron radiation. If the argument of the Bessel functions is small, $\beta \ll 1$, the emission is dominated by low multipoles, $n \simeq 1$ (cyclotron radiation with most of the radiation occurring at $\omega \simeq \omega_B$). As the argument of Bessel functions approaches unity, the
contribution of higher multipoles  increases. In the ultra-relativistic case with $\theta \simeq 0$ (emission close to the plane of rotation), the emission is dominated by multipoles $n \lesssim \gamma^3$, with an exponential cut-off at large frequencies. For $\gamma \gg 1$, the  spectral gap between successive multipoles $\Delta\omega =\omega_B \ll \omega$, and the emission spectrum  
is near-continuum (synchrotron radiation). We discuss the case of non-zero  $\theta$ below.

Eq.~(\ref{eq:synrad_ang}) gives the synchrotron spectrum for a single electron of energy $E = m_e c^2 \gamma$. We consider a range of electron energies and model the electron energy distribution using an exponential cutoff power law $f(\gamma) = N\gamma^{-p} \exp(-\gamma/\gamma_{\rm max})$ in the range $\gamma_{\rm min}$ and $\gamma_{\rm max}$ (e.g., \citealt{Reynolds1999}). $N$ gives the overall normalization. For our work, we treat $\gamma_{\rm min}$ as a free parameter and $\gamma_{\rm max} = 1000 \gamma_{\rm min}$. We use the energy spectral index $p =2.2$, which is consistent with the shock acceleration mechanism (e.g., \citealt{Allen2001}). For this case, if $\gamma_{\rm max}$ is larger than $\gamma_{\rm min}$ by more than a few factors of 10, its impact on our results is found to be negligible.   The factors needed for conversion to flux units for comparison with the data are absorbed in the definition of $N$: $N = \rho_N V /D^2$, where $\rho_N$ is the number density
of relativistic electrons, $V$ is the volume of the emitting region, and $D$ is the luminosity distance to the source.

\begin{figure}
    \centering
    \includegraphics[width=0.45\textwidth]{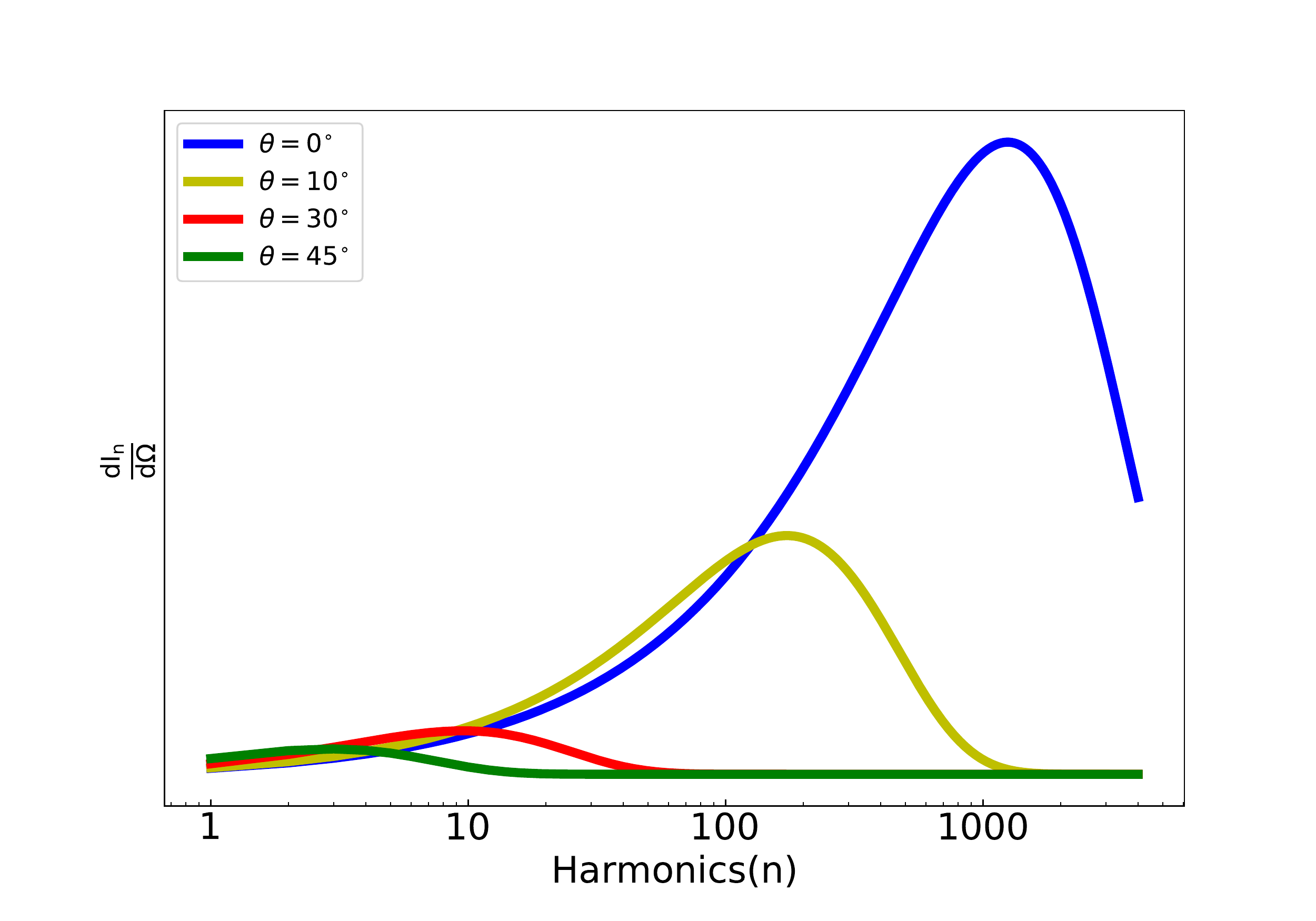}
    \caption{For $B=5 \times 10^{11}$ Gauss and $\gamma=10$, the synchrotron spectra are shown as a function of multipoles  (Eq.~(\ref{eq:synrad_ang}))  for different latitudes. The figure shows the role of high-latitude emission  in  introducing spectral curvature.}
    \label{fig:diff_angle_spectrum}
\end{figure}

As noted above, Eq.~(\ref{eq:synrad_ang}) allows one to analyze the transition from cyclotron to synchrotron radiation. For large $\gamma$, the emission is dominated by large $n$ and is restricted to an angle $\theta \simeq 1/\gamma$ centered on the plane of the orbit. For intermediate 
$\gamma$ or semi-relativistic electrons ($\gamma \lesssim 10$), it is possible to have substantial emissions from higher
latitudes. 
In this paper, we explore the possibility that the observed radiation could emanate from high latitudes with respect to the plane of the orbit.  In this case,  $\beta' = \beta \cos\theta$ acts as the effective velocity parameter in  Eq.~(\ref{eq:synrad_ang}) and determines the frequency at which the synchrotron spectrum begins to fall exponentially.  In Figure~\ref{fig:diff_angle_spectrum}, we
show synchrotron spectra for different values of $\theta$. As expected, for a fixed $\beta$ and $B$,  the spectral cut-off shifts to smaller harmonics $n$ for
larger $\theta$.

Figure~\ref{fig:diff_angle_spectrum} is based on the numerical evaluation of Bessel functions in 
Eq.~(\ref{eq:synrad_ang}). One can gain more direct insight into the
relevant physics with  analytic approximations.  In the literature,  such analytic expressions have been computed for  angle-averaged emission for $\beta \simeq 1$ (e.g., \citealt{Schwinger}). However, such approximations are not valid here as the relevant parameter for us is $\beta \cos\theta$, which can deviate significantly from unity for large angles even for $\beta \simeq 1$. We find that   it is possible to approximate the Bessel function and its derivative in Eq.~(\ref{eq:synrad_ang}) using the stationary phase approximation even when $\beta'$ deviates significantly from unity  (\citealt{Schwinger} employs this method in  the angle-averaged case for $\beta \simeq 1$). This allows us to obtain the following approximate expressions for the Bessel function and its derivative (see Appendix~\ref{sec:AppendixA} for details):
\begin{align}
  J_n(n\beta') & \simeq  0.447 n^{-1/3} \beta'^{-1/3} \> \> {\rm for}\> \> n \le n_c \nonumber \\
  J_n(n\beta') & \simeq 0.335 n^{-1/2} (1-\beta')^{-1/4}\beta'^{-1/4} \nonumber \\
  &\times \exp\left(-2\frac{\sqrt{2}}{3} n \beta'^{-1/2}(1-\beta')^{3/2} \right ) \> \> {\rm for}\> \> n \ge n_c \nonumber \\
   J'_n(n\beta') & \simeq  0.411 n^{-2/3} \beta'^{-2/3} \> \> {\rm for}\> \> n \le n_c \nonumber \\
   J'_n(n\beta') & \simeq 0.474 n^{-1/2} (1-\beta')^{1/4}\beta'^{-3/4}  \nonumber \\
   &\times \exp\left (-2 \frac{\sqrt{2}}{3} n \beta'^{-1/2}(1-\beta')^{3/2} \right ) \> \> {\rm for}\> \> n \ge n_c 
   \label{eq:bess_appro}
\end{align}
with 
\begin{equation}
 n_c \simeq \frac{\beta'^{1/2}}{(1-\beta')^{3/2}}.
\label{eq:cutoffnu}
\end{equation}
$n_c$ denotes the harmonic at which spectral cutoff occurs. The analytic expressions
given in Eq.~(\ref{eq:bess_appro}) agree with numerical results to better
than 10\% in the acceptable range of $\beta' \gtrsim 0.3$.  Also, for 
$\beta' \simeq 1$ ($\beta \simeq 1$ and $\theta \simeq 0$), $1/(1-\beta') \simeq 2\gamma^2$, and  $n_c \simeq \gamma^3$ which agrees with the angle-averaged case  (\citealt{Schwinger}).

Our aim in this paper is to explain  spectral cutoff as observed in ULXs. For
fitting the  X-ray continuum spectral data, we require $n_c \gg 1$, which constrains the latitude $\theta \lesssim 70^{\circ}$. One can consider the intriguing
possibility that the observed spectrum could arise from a set of discrete lines
(though it is unlikely, as we argue below). This requires: 
$n_c \ge 1$, which gives $0.3 \lesssim \beta \lesssim 1$ and $0.3 \lesssim \cos\theta \lesssim 1$ \footnote{\url{https://www.wolfram.com/mathematica/}}. We note that the exponential terms in Eq.~(\ref{eq:bess_appro}) adequately capture the cut-off frequency in the entire parameter range of interest, which is key to modeling the ULX cut-off frequency.   While we compute Bessel functions numerically for data analysis, these analytic expressions help us interpret our results.

In Figure~\ref{fig:diff_angle_spectrum}, we display synchrotron spectra for emission from different latitudes. Eq.~(\ref{eq:bess_appro}) allows us to understand the spectral shapes seen in the figure. The spectral cut-off occurs
at an angular frequency $\omega \simeq n_c \omega_B$. For emission close to the plane of the rotation ($\theta \simeq 0$), $n_c \simeq \gamma^3$. However, for larger angles $n_c < \gamma^3$, as Eq.~(\ref{eq:bess_appro}) shows, and the spectral cut-off shifts to smaller frequencies. As we discuss later, the spectral cut-off in the data we analyze occurs at $E \simeq 10~\, \rm keV$, which is possible for a range of $\gamma$, $B$, and $\theta$ as will be discussed below in more detail. Even though we only assume electron motion
in the plane perpendicular to the magnetic field, our results do not qualitatively change if the electron has a $z$-component of velocity. This case can be incorporated into our analysis by altering $B$ to $B_\perp \equiv B \cos\chi$, where $\chi$ is the angle between the velocity vector and the magnetic field (e.g., \citealt{Landau}).

\section{Data} \label{sec:data}

We utilize the \nustar\ observations of two bright pulsar ULXs, NGC 5907 ULX1 (RA:15 15 58.62, DEC: +56 18 10.3; \citealt{Israel2017feb,Walton2015,Furst2017}) and NGC 7793 P13 (RA: 23 57 50.9, DEC: -32 37 26.6; \citealt{Furst2016,Israel2017mar,Walton2018feb}) to compare  our  theoretical model against  data. The distances to the host galaxies are $\simeq 17.1$ Mpc (e.g.,  \citealt{Furst2017}) and $\simeq 3.5$~Mpc (e.g., \citealt{Walton2018feb}),  respectively. These two sources were observed by \nustar\ several times in the past decade, which provide us an opportunity to verify the consistency of the theory over long-term spectral evolution of the sources.  In particular, for NGC 7793 P13, we detect two distinct flux states. The choice of the instrument is motivated by its energy coverage, which allows us to model the spectral curvature of the source. The broadband spectra of ULXs are generally fitted with multiple components: neutral absorption, accretion disk (geometrically thin or slim), and a phenomenological model of either magnetic or non-magnetic Comptonization processes (see e.g., \citealt{Kaaret2017}). The thermal disk component and neutral absorption mostly play a role in the soft energy regime ($E \lesssim 5 \, \rm keV$). Our aim in this paper is to explain the spectral cutoff in ULXs, which occurs in a higher energy range ($E \simeq \rm 10 keV$). To minimize contamination from soft components and to adequately model the spectral break, we study  the energy  range $\simeq 5\hbox{--}25$~keV in this paper.  In our study, we consider all the available \nustar\ data sets for both the sources. However, for NGC 5907 ULX1, there are a few observations for which  the  signal-to-noise ratio (S/N) is poor owing to the faintness of the source. We do not utilize these data for our analysis. 

\subsection{Data reduction process}\label{subsec:data_reduction}
The \nustar\ data are extracted using the \texttt{HEASOFT} version 6.29 \footnote{\url{https://heasarc.gsfc.nasa.gov/docs/software/heasoft/}}. We use \texttt{nupipeline} tool to extract cleaned products and \texttt{nuproducts} tool to extract the source and background spectra and the response files from both FPMA and FPMB modules. In general we follow the method outlined in previous works (e.g., \citealt{Israel2017feb,Walton2015,Furst2017,Furst2016,Israel2017mar,Walton2018feb,Lin2022}) for data reduction of these two sources. We choose the source photon extraction region as $50$ arcsec radius circle for both sources. The background regions are selected as $100$ arcsec radius circle in all the cases. The number of counts per energy bin for grouping the spectra are a minimum of $30$ counts per energy bin for all NGC 5907 ULX1 spectra and $50$ counts per bin for NGC 7793 P13 spectra where the source is in high flux state, and $20$ counts per energy bin for low flux state (observation IDs - 30502019002, 30502019004, 50401003002, 90601327002) of the source.

After we obtain the spectra, we use \texttt{XSPEC} \citep{XSPEC} spectral analysis package to convert the spectra into flux units for further analysis. \nustar\ spectra beyond $\simeq 25$~keV are dominated by the background for both the sources, and therefore spectral data above this energy are not utilized in our analysis. In some low flux state observations, the background starts dominating well below $\simeq 20$ keV; however, to provide similar treatment to all observations, we take spectra up to $\simeq 25$~keV for all cases. We fit the \nustar\ spectra with a cutoff power-law model (in \texttt{XSPEC} the syntax is \texttt{constant*cutoffpl}). The \texttt{constant} model represents the instrumental cross-calibration differences, and \texttt{cutoffpl} is the continuum representing an exponentially cutoff power-law spectrum. This model for $5.0\hbox{--}25.0$~keV spectra give statistically good fit for both the sources. For the cutoffpl model, we fix the index to $0.59$, a typical value for ULX pulsars (see e.g., \citealt{Walton2020}). We do not consider neutral absorption, since it plays a role  only in softer regime of the spectra. We then convert the spectral counts into flux $\rm \nu F_\nu$ (ergs/cm$^2$/sec) by the \texttt{eeufspec} command and take the data points in the energy range $5.0\hbox{--}25.0$~keV ($\simeq 1.2\hbox{--}6.0 \times 10^{18}$~Hz) to perform further analysis described in the \S~\ref{sec:Results}. We have further verified the robustness of this data extraction procedure with another model such as a simple \texttt{powerlaw} of photon index $0$, instead of the \texttt{cutoffpl} model, and extract the spectra in flux unit using \texttt{eeufspec}. We find that our results are insensitive to the  choice of \texttt{XSPEC} models used to generate the flux data points.

\section{Analysis and Results} \label{sec:Results}

From Eq.~\ref{eq:synrad_ang}, one can verify that $\gamma$ and $B$ are degenerate with each other if $\beta \simeq 1$. As we wish to explore a range of electron speeds from semi-relativistic to ultra-relativistic, this degeneracy cannot be removed. {\footnote{$\beta \simeq 1$ approximation is appropriate for ultra-relativistic case or semi-relativistic case in higher latitudes.}} Thus we choose $\gamma_{\rm min}/B$ as one of the parameters in the analysis. We use three parameters---$\gamma_{\rm min}/B$, $\theta$ and $N$---in our analysis and later fix one more parameter to deal with residual degeneracies. We also convert the model to $\rm \nu F_\nu$ unit by appropriately scaling Eq.~\ref{eq:synrad_ang} by a multiplicative factor $n = \omega/\omega_B$, where $\omega = 2\pi\nu$.

For our model, the observed flux can be written  as:
\begin{align}
  \nu F_\nu = N'
  \int_{x_{\rm min}}^{x_{\rm max}} S x^{-p} \exp\left(-x/x_{\rm max}\right) dx
  \label{eq:total_spec}
\end{align}
where $x =\gamma/B$,
\begin{equation}
 N' = \frac{N}{\int_{x_{\rm min}}^{x_{\rm max}} x^{-p} \exp\left(-x/x_{\rm max}\right) dx}
\end{equation}
and,
\begin{equation}
 S = \frac{2\pi e^2\nu^3}{c\nu_B} \left[\tan^2\theta J^2_\frac{\nu}{\nu_B}\left(\frac{\nu}{\nu_B}\cos\theta\right)  + J'^2_\frac{\nu}{\nu_B}\left(\frac{\nu}{\nu_B}\cos\theta\right) \right]
\end{equation}

\begin{figure*}[t]
    \centering
    \hspace{-2cm}
    \includegraphics[width=0.55\linewidth]{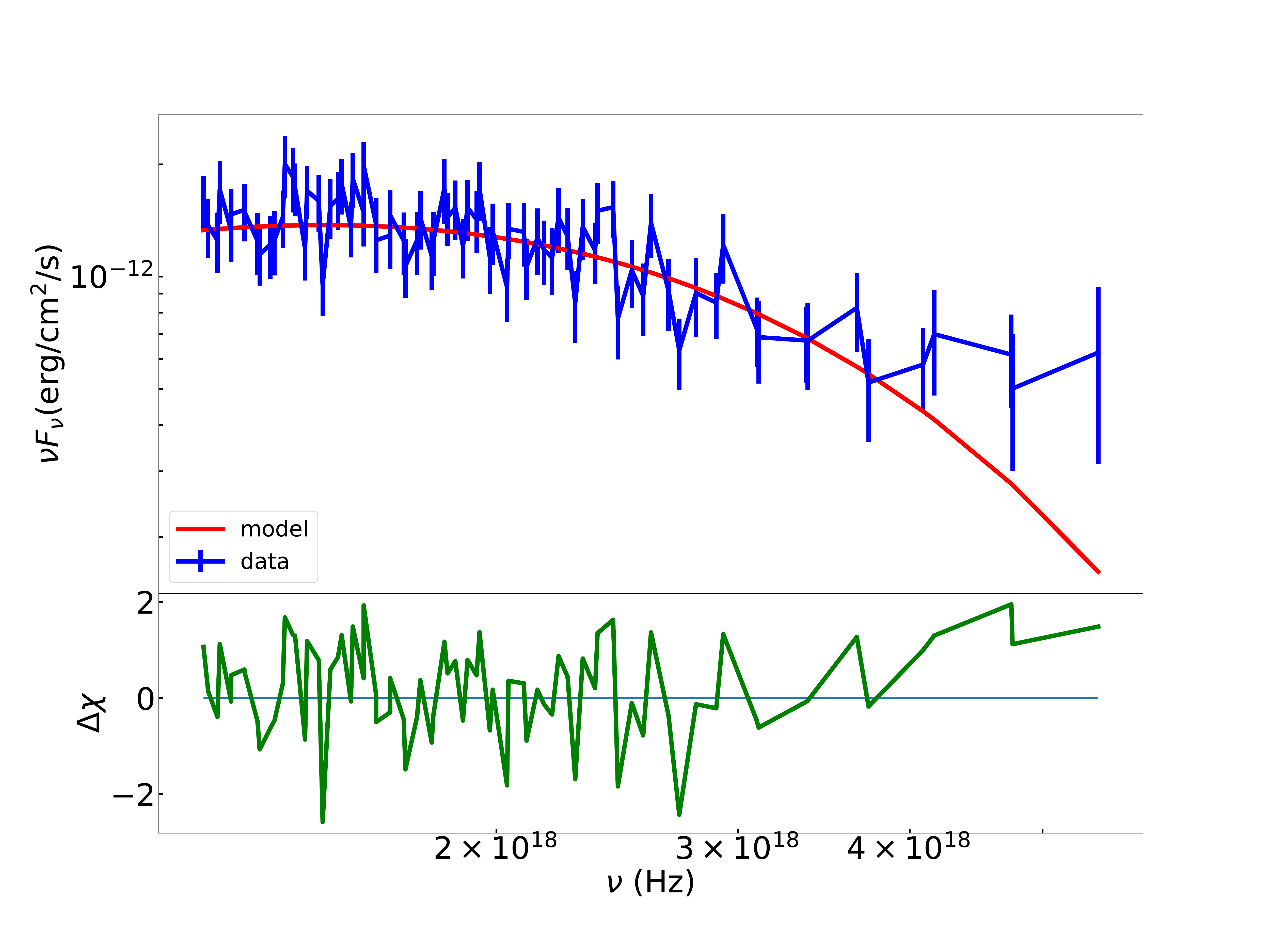}
    \hspace{-1cm}
    \includegraphics[width=0.55\linewidth]{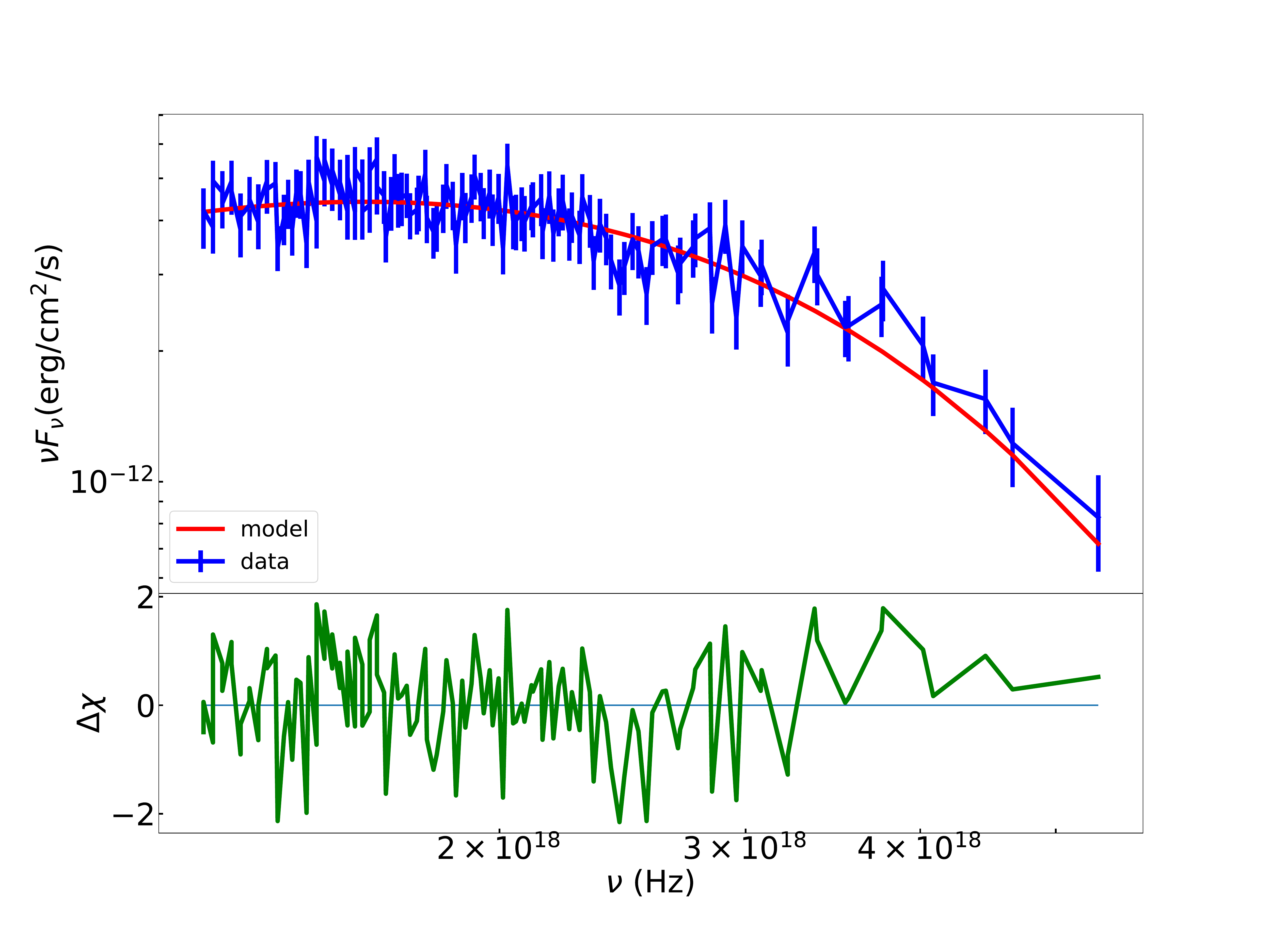}
    \hspace{-2cm}
    \caption{Example plot showing spectral fitting of the theoretical model on one observation for each source. The top panels show the data with model over plotted and corresponding residuals are shown in bottom panels. The model shown here are for $1^{\circ}$ angle in case of NGC 5907 ULX1 (left; Observation - 80001042002) and $30^{\circ}$ angle for NGC 7793 P13 (right; Observation - 90301326002).}
    \label{fig:residual_plots}
\end{figure*}

To explore the congruence of the data and the model, we choose two different statistical methods-- frequentist approach and Bayesian analysis. Given that it is hard to determine the best-fit and the errors on the three parameters simultaneously, we fix the angle $\theta$ and keep the other two parameters $\gamma_{\rm min}/B=x_{\rm min}$ and $N$ free to vary. For the frequentist approach, we first carry out a minimum $\chi^2$ analysis using the \texttt{scipy} \citep{2020SciPy-NMeth} ``curve\_fit" tool \footnote{\url{https://docs.scipy.org/doc/scipy/reference/generated/scipy.optimize.curve_fit.html}}. The best-fit parameters, $\chi^2$ values, and 1$\sigma$ errors (computed using covariance matrix) are given in Table~\ref{tab:parameters}. For each data set, we consider three values of $\theta = 1^{\circ}, 15^{\circ}, 30^{\circ}$ for our analysis.  As discussed in the foregoing, this choice is based on the acceptable range of latitudes to ensure $n_c \gg 1$. In Figure~\ref{fig:residual_plots} we plot the data, the bestfit curve and   spectral residuals for one observation for each source.

In the Bayesian analysis, we utilize the Markov chain Monte Carlo (MCMC) method using  \texttt{python} \texttt{emcee} package \footnote{\url{https://emcee.readthedocs.io/en/stable/}} \citep{emcee}. We find convergence in each case and the computed posterior probabilities agree with the results obtained using the frequentist method.

We next discuss the physical implications of the parameter range suggested by statistical analyses. We find that for fixed emission angles, the spectrum is scaled by the parameter $x_{\rm min}$ and the overall amplitude is scaled by $N$. For both the sources, we do not find significant long-term spectral variability in the high-energy band as the parameter $x_{\rm min}$ remains nearly the same for all observations (Table~\ref{tab:parameters}). However, due to flux changes in different epochs of observation, the parameter $N$ varies significantly.

For the range of $\theta$ shown in Table~\ref{tab:parameters}, the estimated range of  
$x_{\rm min} = \gamma_{\rm min}/B$ 
varies from $10^{-6} \, \rm G^{-1}$ to $10^{-11} \, \rm G^{-1}$. It can be  verified from the expression for $\nu_B = \omega_B/(2\pi)$ and Eq.~(\ref{eq:cutoffnu}) that the results given in Table~\ref{tab:parameters} correspond
to a cutoff frequency of around 10~keV. The allowed range of $x_{\rm  min}$ encompasses a large range of 
particle speeds and magnetic field strengths.  At higher latitudes, our results are consistent with  semi-relativistic electrons, $\gamma_{\rm min} \simeq 20$ and $B \simeq 10^{12} \, \rm G$. This magnetic field strength is expected on the surface of neutron stars (e.g., \citealt{Caballero2012, Petri2016}). The overall normalization $N = \rho_N V/D^2$ is highly uncertain as both the relativistic electron density $\rho_N$ and the volume of emission region $V$ are very poorly determined even theoretically. Typically, in neutron star magnetosphere, the lower limit of the plasma density is given by the Goldreich-Julian limit \citep{Goldreich} which depends on pulsar spin period, magnetic field strength, and alignment of spinning axis with magnetic field line. Depending on the volume of emission region, we find that the estimated number density could be comparable to or higher than the Goldreich-Julian limit \citep{Goldreich} for a $1$ sec spinning pulsar with $B \simeq 10^{12}$~G , i.e., $7 \times 10^{10}$ particles $\rm cm^{-3}$. In reality, the plasma density can be significantly higher than the Goldreich-Julian limit (see \citealt{Lyutikov} and references therein). This means our results are consistent with this theoretical expectation. We also determine  that, for a range of acceptable parameters, the emitting region is optically thin to synchrotron self-absorption and compton scattering.

If we restrict the maximum limit of magnetic field on the NS surface to the Schwinger limit of $4.4 \times 10^{13}$~G, the magnetic field strength at which the quantum effects become important,  then the maximum value of $\gamma_{\rm min}$ can be estimated. For lower $x_{\rm min}$ (i.e., lower $\gamma_{\rm min}$ or higher $B$), we get emission at higher latitudes which requires lower value of $N$ to explain the observed ULX flux. On the other hand, when $x_{\rm min}$ is higher, we get emission closer to the plane of orbit and higher $N$ is required to generate such high flux in these sources. It would be possible in the future to constrain all the parameters adequately if we can have at least one parameter determined  from other data. Our results point to the possibility that the spectral curvature in ULXs might have a common origin and all the ULXs are possibly highly magnetized neutron stars. This theoretical model can also be employed to explain  the high energy cutoff in  X-ray binary pulsar sources. Essentially, this model suggest that the curvature in the spectrum is governed by plasma velocity, magnetic field strength, electron number density, and emission angle.

\begin{table*}
\scriptsize
\centering
\caption{The best fit parameters and $\chi^2$ from python curve fit \label{tab:parameters} for 7 \nustar\ observations of NGC 5907 ULX1 and 10 \nustar\ observations of NGC 7793 P13 are listed. The errors are calculated for  1-$\sigma$ confidence from the covariance matrix using the parameter \texttt{absoulte\_sigma=True}.}
\begin{tabular}{CCCC|CCC|CCC}
\hline
\noalign{\smallskip}
&\multicolumn{3}{c|}{$\theta = 1^{\circ}$}&\multicolumn{3}{c|}{$\theta = 15^{\circ}$}&\multicolumn{3}{c|}{$\theta = 30^{\circ}$}\\
\hline 
$\rm Observations$ & \rm x_{min}  &  \rm N & \rm \chi^2 /d.o.f & \rm x_{min}  &  \rm N  & \rm \chi^2 /d.o.f  & \rm x_{min}   &  \rm N  & \rm \chi^2 /d.o.f   \\
 \hline
 & (10^{-7} ~G^{-1})  & (10^{-17} ~cm^{-2}) &  & (10^{-10} ~G^{-1})  & (10^{-18} ~cm^{-2}) &    & (10^{-11} ~G^{-1})  & (10^{-19} ~cm^{-2}) &    \\
\hline
\noalign{\smallskip}
\textrm{NGC 5907 ULX1}\\
\hline
30002039005 & (9.54 \pm 0.56) & (1.10 \pm 0.18) & 57/49  & (2.81 \pm 0.16) & (0.70 \pm 0.12) & 57/49  & (3.45 \pm 0.20) & (2.97 \pm 0.49) & 57/49 \\
30302004006 & (6.38 \pm 0.36) & (1.00 \pm 0.12) & 66/52 &(1.88 \pm 0.11)  & (0.63 \pm 0.08) & 66/52 & (2.31 \pm 0.13) & (2.72 \pm 0.34) & 66/52 \\
30302004008 & (5.94 \pm 0.35) & (0.81 \pm 0.10) & 62/49 &(1.75 \pm 0.10) & (0.51 \pm 0.06) & 62/49 & (2.15 \pm 0.12) & (2.21 \pm 0.27) & 62/49 \\
80001042002 & (7.69 \pm 0.27)  & (2.52 \pm 0.23) & 80/78 & (2.27 \pm 0.08)  & (1.60 \pm 0.14)  & 80/78 & (2.78 \pm 0.10)  & (6.84 \pm 0.61)  & 80/78 \\
80001042004 & (6.57 \pm 0.26)  & (1.69 \pm 0.15) & 82/77 & (1.94 \pm 0.08) & (1.07 \pm 0.10)  & 82/77 & (2.38 \pm 0.09) & (4.58 \pm 0.41)  & 83/77\\
90501331002 & (5.97 \pm 0.46) & (0.47 \pm 0.08) & 34/39 & (1.76 \pm 0.14)  & (0.30 \pm 0.05)  & 34/39 & (2.16 \pm 0.16)  & (1.28 \pm 0.21) & 34/39 \\
90601323002 & (7.23 \pm 0.29) & (1.66 \pm 0.16)  & 89/88 & (2.13 \pm 0.09)  & (1.05 \pm 0.10)  & 90/88 & (2.62 \pm 0.10)  & (4.50 \pm 0.43) & 90/88\\
\hline
\noalign{\smallskip}
\textrm{NGC 7793 P13}\\
\hline
30302005002 & (7.66 \pm 0.20)  &  (3.07 \pm 0.20)  & 77/82 & (2.26 \pm 0.06) & (1.95 \pm 0.13)  & 77/82 & (2.77 \pm 0.07) & (8.30 \pm 0.55) & 77/82 \\
30302005004 & (7.46 \pm 0.15) & (4.58 \pm 0.23)  & 111/117 & (2.20 \pm 0.04)  & (2.90 \pm 0.15)  & 111/117 & (2.70 \pm 0.05)  & (12.40 \pm 0.62) & 111/117 \\
30302015002 &  (7.53 \pm 0.14)  & (5.95 \pm 0.27)  & 147/133 & (2.22 \pm 0.04)  & (3.77 \pm 0.17)  & 147/133 & (2.73 \pm 0.05)  & (16.10 \pm 0.73)  & 148/133 \\
30302015004 & (7.53 \pm 0.17)  & (4.75 \pm 0.26)  & 130/123 & (2.22 \pm 0.05)  & (3.02 \pm 0.17)  & 130/123 & (2.72 \pm 0.06)  & (12.88 \pm 0.70)  & 130/123 \\
30502019002 & (7.41 \pm 0.53)  & (0.65 \pm 0.11) & 42/61 & (2.18 \pm 0.16)  & (0.41 \pm 0.07) & 42/61 & (2.68 \pm 0.19)  & (1.77 \pm 0.30)  & 42/61 \\
30502019004 & (6.93 \pm 0.62)  & (0.57 \pm 0.12)  & 32/42 & (2.04 \pm 0.18) & (0.36 \pm 0.08)  & 32/42 & (2.51 \pm 0.22)  & (1.55 \pm 0.32)  & 32/42\\
50401003002 & (7.37 \pm 0.79) & (0.59 \pm 0.15)  & 25/31 & (2.17 \pm 0.23)  & (0.38 \pm 0.10)  & 25/31 & (2.67 \pm 0.28)  & (1.60 \pm 0.41)  & 25/31\\
80201010002 & (7.26 \pm 0.10)  & (6.20 \pm 0.22)  & 156/204 & (2.14 \pm 0.03)  & (3.94 \pm 0.14)  & 156/204 & (2.63 \pm 0.04)  & (16.81 \pm 0.58) & 156/204\\
90301326002 & (7.20 \pm 0.14)  & (7.11 \pm 0.33)  & 104/123 & (2.12 \pm 0.04) & (4.51 \pm 0.21) & 104/123 & (2.61 \pm 0.05) & (19.26 \pm 0.89)  & 105/123\\
90601327002 & (9.65 \pm 1.01) & (1.00 \pm 0.30)  & 35/34 & (2.84 \pm 0.30)  & (0.63 \pm 0.19)  & 35/34 & (3.49 \pm 0.36)  & (2.70 \pm 0.80) & 35/34\\
\hline
\end{tabular}
\end{table*}

\section{Discussion}\label{sec:discussion}
In this paper, we seek to explain the X-ray spectra of two known pulsar ULXs. In particular, we focus on the  cut-off in such spectra at $E \simeq 10$~keV, which is a generic feature of  many pulsar ULXs.
We propose synchrotron radiation at a range of latitudes  as the possible physical process to explain this observed spectral shape. Our main results, based on the analysis  of 17  spectra of the two sources, are summarized in Table~\ref{tab:parameters}. Figure~\ref{fig:residual_plots} shows the fit and its residual for one spectrum for each source.

Other  models that have been explored  to explain the dominant emission in hard  X-ray range in ULXs invoke the comptonization from coronal region of a non-magnetic sources and comptonization from magnetized column in neutron stars (see e.g., \citealt{West2018, Walton2020}). Compton scattering in presence of high magnetic field in neutron star is a possible candidate to explain high luminosity in these sources \citep{Mushtukov2015}. In principle, there could be two possible ways to distinguish our proposed scenario from these models.

Table~\ref{tab:parameters} shows that the best-fit value $x_{\rm min} = \gamma_{\rm min}/B \simeq 2 \times 10^{-11}$ for $\theta \simeq 30^\circ$. This could correspond, for instance, to a semi-relativistic electron ($\gamma_{\rm min} \simeq 20$) along with a magnetic field $B\simeq 10^{12} \, \rm G$. In such cases, the fundamental mode of emission, $\nu_B = eB/(2\pi m_ec\gamma_{\rm min}) \simeq  600\, \rm eV$. As this is larger than the spectral resolution of \nustar\ in energy range of interest, the observed spectrum could be a set of discrete cyclotron lines. In practice, such an interpretation could be difficult owing to mixing with larger $\gamma$ values and the width of spectral lines, which are  difficult to ascertain. This would also require either re-analysis of the data or new data which is beyond the scope of the paper. Our analysis raises the intriguing possibility that the discreteness of the spectrum could probe the latitude of the emission. As we have already discussed in \S~\ref{sec:Theory}, an  upper limit on the  latitude  of emission can be obtained by  requiring $n_c \gg 1$; this yields a stringent upper bound $\theta \simeq 70{^\circ}$. For fitting  continuum X-ray spectral data  in the energy range of interest,  this requirement    motivates the upper limit of   $\theta \simeq 30^{\circ}$ we use in this paper.

Another possible probe of our model could be the polarization of received photons. The photons emerging from higher latitudes would be elliptically polarized while those from closer to the plane of the orbit would be linearly polarized. While the non-magnetic comptonization will not show polarized emission, the magnetic comptonization and the synchrotron radiation could display different degrees of polarization. Modern X-ray polarimeters such as IXPE \citep{Ixpe} and upcoming mission POLIX \citep{Polix} might be able to address these questions.

We provide a brief summary of our main results and perspectives below:
\begin{itemize}
\item[1.] We propose that spectral cutoff in  ULXs arise from  classical, high-latitude, and  optically-thin synchrotron radiation. For classical radiation, the cutoff occurs at energies $\gamma^3 \nu_B$ for radiation close to the plane of the orbit, but the cutoff frequency shifts to much smaller frequencies for high-latitude emission. Quantum effects only dominate the cutoff for energies close to the electron rest mass  and hence cannot be responsible for the observed  cutoff at $E \simeq 10~\, \rm keV$.

\item[2.] This model is compared with the 17  spectra, corresponding to different flux states of two ULXs. The observed fluxes are  modeled using four theoretical parameters. Given the degeneracy between these parameters, only two parameters can be  estimated from the data.  To test  the robustness of our statistical analysis, we  carry out both frequentist and bayesian analysis (using MCMC). While our analysis yields a large range of possible theoretical models, the most interesting case
corresponds to high-latitude  emission  ($\theta \simeq 30^\circ$) from a  semi-relativistic plasma from the surface of  the neutron star ( $B\simeq 10^{12} \, \rm G$ and $\gamma \simeq 20$). The statistical analysis also allows us
to establish that the plasma is optically thin for a plausible range of parameters.

\item[3.] It is possible to verify the model using X-ray polarization data which might be available in the near future. Another possible probe of the semi-relativistic plasma could be the discreteness of the observed spectrum, which we have briefly discussed.

\end{itemize}

\begin{acknowledgments}
We would like to thank the referee for the valuable suggestions that helped further improve the manuscript. We would like to thank Keith Arnaud from HEASARC helpdesk for his valuable suggestions regarding eeufspec tool in XSPEC. This research has utilized archival data (available at the High Energy Astrophysics Science Archive Research Center (HEASARC)) obtained with \nustar, a project led by Caltech, funded by NASA, and managed by the NASA Jet Propulsion Laboratory (JPL), and has made use of the NuSTAR Data Analysis Software (\texttt{NuSTARDAS}) jointly developed by the ASI Space Science Data Centre (SSDC, Italy) and the California Institute of Technology (Caltech, USA).
\end{acknowledgments}

%

\vspace{5mm}
\facilities{\nustar ; \citealt{NuSTAR} }


\software{HEASOFT (\url{https://heasarc.gsfc.nasa.gov/docs/software/heasoft/}; \citet{Heasoft2014}), Mathematica (\url{https://www.wolfram.com/mathematica/}; \citet{Mathematica})
          }





\appendix

\section{Analytic approximation of Bessel function}\label{sec:AppendixA}
The starting point of approximating the Bessel function and its derivatives is
the integral representation of these functions (e.g. \cite{Schwinger,Landau}),
\begin{align}
    J_n(z) = \int_0^\pi d\phi \frac{1}{\pi} \cos(z\sin\phi-n\phi)\\
    J'_n(z) = -\int_0^\pi d\phi \frac{1}{\pi} \sin\phi \sin(z\sin\phi-n\phi)
\end{align}
Here $z = n\beta\cos\theta = n \beta'$. As the integrads are highly oscillatory, the main contribution to the integrals arise from  regions near $\phi=0$ when the phase is large (stationary phase approximation). This is ensured by the condition $n\beta' \gg 1$. Expanding the phase  of $J_n(z)$ around $\phi = 0$, we get:
\begin{equation}
    z\sin\phi-n\phi = n \beta'\sin\phi-n\phi = - n \left[\phi(1-\beta')+\frac{\beta'\phi^3}{3!}\right ]
\end{equation}
Making the substitution,  $\phi = (1-\beta')^{1/2}x/\beta'^{1/2} $ yields:
\begin{equation}
    \phi(1-\beta')+\frac{\beta'\phi^3}{3!} = \frac{(1-\beta')^{3/2}}{\beta'^{1/2}} \left (x+\frac{x^3}{6}\right )
\end{equation}
In this case, the stationary phase points are located at:
\begin{equation}
    x = \pm \sqrt{2}i
\end{equation}
Following the procedure outlined in \citealt{Schwinger},  in the neighborhood of the stationary phase point, we can write:
\begin{equation}
    x = \sqrt{2}i+\xi,
\end{equation}
where $\xi$ is real and  small, which gives:  
\begin{equation}
    x+\frac{x^3}{6} = \sqrt{2} i \left(\frac{2}{3}+\frac{\xi^2}{2}\right)
\end{equation}
This allows us to write:
\begin{equation}
  J_n(n\beta') = \int_0^\infty dx \frac{1}{\pi} \frac{(1-\beta')^{1/2}}{\beta'^{1/2}} \cos\left(n\left[\frac{(1-\beta')^{3/2}}{\beta'^{1/2}} \left(x+\frac{x^3}{6}\right)\right]\right)
  \label{eq:bessaprro1}
\end{equation}
First, we deal with the case 
when $n (1-\beta')^{3/2}/\beta'^{1/2} \ll  1$. In this case,  the main contribution to the integral  come from the region where $x$ is large.
Given that most of the contribution to the integral comes from regions,
where the phase is close to unity, the integration limit can be extendend to  infinity (e.g. \citep{Schwinger}). Solving the resultant integral,  we get:
\begin{equation}
    J_n(n\beta')  \simeq  0.447 n^{-1/3} \beta'^{-1/3} \> \> {\rm for}\> \> n \frac{(1-\beta')^{3/2}}{\beta'^{1/2}} \ll 1
\end{equation}
For $n (1-\beta')^{3/2}/\beta'^{1/2} \gg  1$,  the integral can be written as:
\begin{equation}
  J_n(n\beta') = Re \int_0^\infty dx \frac{1}{\pi} \frac{(1-\beta')^{1/2}}{\beta'^{1/2}} \exp\left(i n(1-\beta')^{3/2} (x+x^3/6)/\beta'^{1/2}\right)
\end{equation}
This can readily be integrated as most of contribution arises from regions
close to $x\simeq 0$:
\begin{equation}
       J_n(n\beta') \simeq 0.335 n^{-1/2} (1-\beta')^{-1/4}\beta'^{-1/4} \times \exp\left(-2\frac{\sqrt{2}}{3} n \beta'^{-1/2}(1-\beta')^{3/2} \right )
\end{equation}
Following a similar calculation procedure, we obtain the approximate forms of $J'_n(n\beta')$ in Eq. \ref{eq:bess_appro}.

Our analysis extends the procedure outlined by \citealt{Schwinger} for the extreme relativistic case $\beta' \simeq 1$ to  arbitrary $\beta'$.  We note that the analytic expressions we derive yield excellent fits to  numerical results for $\beta' \gtrsim 0.3$.


\bibliography{Synchrotron_cutoff_ApJ}{}
\bibliographystyle{aasjournal}



\end{document}